\documentclass[twocolumn,english,aps,prb,floats]{revtex4}
\usepackage[T1]{fontenc}
\usepackage[latin9]{inputenc}
\usepackage{babel}
\usepackage{bm}
\usepackage{amsmath}
\usepackage{amssymb}
\usepackage{graphicx}
\usepackage{wasysym}
\usepackage{esint}
\usepackage{float}
\usepackage[unicode=true,
 bookmarks=false,
 breaklinks=false,pdfborder={0 0 1},backref=section,colorlinks=false]
 {hyperref}
\usepackage{breakurl}

\makeatletter
\@ifundefined{textcolor}{}
{%
 \definecolor{BLACK}{gray}{0}
 \definecolor{WHITE}{gray}{1}
 \definecolor{RED}{rgb}{1,0,0}
 \definecolor{GREEN}{rgb}{0,1,0}
 \definecolor{BLUE}{rgb}{0,0,1}
 \definecolor{CYAN}{cmyk}{1,0,0,0}
 \definecolor{MAGENTA}{cmyk}{0,1,0,0}
 \definecolor{YELLOW}{cmyk}{0,0,1,0}
}

\@ifundefined{textcolor}{}{%
 \definecolor{BLACK}{gray}{0}
 \definecolor{WHITE}{gray}{1}
 \definecolor{RED}{rgb}{1,0,0}
 \definecolor{GREEN}{rgb}{0,1,0}
 \definecolor{BLUE}{rgb}{0,0,1}
 \definecolor{CYAN}{cmyk}{1,0,0,0}
 \definecolor{MAGENTA}{cmyk}{0,1,0,0}
 \definecolor{YELLOW}{cmyk}{0,0,1,0}
}

\usepackage{babel}
\usepackage{ifpdf}\usepackage{bm}

\makeatother

\begin{document}

\title{Skyrmions in an oblique field: A path to novel binary and quaternary memory}

\author{Daniel Capic, Dmitry A. Garanin, and Eugene M. Chudnovsky}

\affiliation{Physics Department, Herbert H. Lehman College and Graduate School,
The City University of New York, 250 Bedford Park Boulevard West,
Bronx, New York 10468-1589, USA }

\date{\today}
\begin{abstract}
We propose a novel binary and quaternary memory device based upon skyrmion states induced by the oblique field in a square magnetic island. To describe stable states and dynamics of the skyrmion, we employ the lattice model that uses the parameters of a real material and accounts for all relevant interactions. Depending on the orientation of the field, two or four spatially separated energy minima emerge in the oblique field. The energy barriers between the minima can be controlled  by the strength and orientation of the magnetic field. We study the dynamics of the skyrmion and show that it can be moved between any two states by application of the field gradient. Islands of thickness of a few tens of atomic layers permit room-temperature manipulation of the proposed device. 

\end{abstract}

\pacs{}
\maketitle

\section{Introduction}

Topological properties of skyrmions and their potential for applications (see, e.g., Zhang \textit{et al}., Ref.\ \onlinecite{Zhang2020}) have brought skyrmions to the forefront of research in magnetism. Writing and reading information with skyrmions have attracted particularly large attention in recent years. Logic gates that involve the conversion of skyrmions into domain walls \cite{ Zhou2014,  Xing2016, ZhangLogic2015, ZhangLogicTwo2015} have been studied in detail. It was shown in Ref. [\onlinecite{Zhang2015}] that a  skyrmion-based transistor device can be constructed by the application of a local electric field that changes the perpendicular magnetic anisotropy (PMA) of the system. This creates an energy barrier that prevents the skyrmion from passing the voltage-gated region, thus creating the OFF state. In the ON state the skyrmion is driven past the gated region with the spin current.  A similar idea of a transistor-like device was also explored by Upadhyaya \textit{et al} \cite{Upadhyaya2015}. Xia \textit{et al}.\cite{Xia2017} expanded upon this by driving skyrmions with a microwave field. Most recently, Zhang \textit{et al}. \cite{Zhang2020Logic} proposed the use of skyrmions as elements of logic for stochastic computing, where the skyrmion can be controlled by changing the PMA or the Dzyaloshinskii-Moriya interaction. Furthermore, Wang \textit{et al}. \cite{Wang2017Gradient} have shown that it is possible to manipulate and trap skyrmions with a magnetic field gradient. Other methods of skyrmion manipulation are possible, but not explored in this work.

The repulsion of the skyrmion as it approaches the boundary is well-established. Rohart and Thiaville \cite{Rohart2013} computed the bending of the magnetization at the boundary by the Dzyaloshinskii-Moriya (DM) interaction which can lead to the confinement of skyrmions. Sanchez \textit{et al}. \cite{Sanchez2014} expanded on this work to obtain explicit twisted boundary conditions for the Landau-Lifshitz equation when the DMI is present.
  
One of the earliest practical works on the skyrmion dynamics in a confined geometry was done by Iwasaki \textit{et al} \cite{Iwasaki2013} within the rigid Thiele body formalism where the repulsion with the boundary is introduced as a potential $V$. They found that the skyrmion velocity as a function of current for the steady state is similar to domain walls.  Zhang \textit{et al}. \cite{ZhangBoundary2015} studied skyrmion repulsion in a skyrmion-based racetrack memory, one feature of which is that the equilibrium distance of the skyrmion from the boundary depends on the track width. Leonov and Mostovoy \cite{Leonov2017Edge} studied skyrmion dynamics with the edges of frustrated magnets that produce multiple edge channels that affect the motion of the skyrmion. Yoo \textit{et al}.\cite{Yoo2017} saw that the threshold current density to expel the skyrmion depended on the critical repulsive boundary force. Martinez \textit{et al}. \cite{Martinez2018} expanded upon this to study the skyrmion dynamics near the boundary and saw an asymmetry in the motion that depended on the damping parameters, which in turn are related to the anisotropy energy. Menezes \textit{et al} \cite{Menezes2019} observe the deflection of (ferromagnetic) skyrmions at a heterochiral interface that is more pronounced when the difference in DMI across the interface is larger.  Jin \textit{et al}. \cite{Jin2020} use the skyrmion confinement by an annular groove to produce a frequency tunability that is much higher than similiar skyrmion based spin transfer nano-oscillators (STNOs). Most recently, Brearton \textit{et al}. \cite{Brearton2020} found that the repulsion by the boundary can be modeled well as a decaying exponential function.

Additionally, a number of works have studied skyrmions and skyrmion lattices in the presence of an oblique field. Lin and Saxena \cite{Lin2015} studied the equilibrium and dynamical properties of skyrmions in thin films of chiral magnets with a tilted field and observed the elongation of the Bloch-type skyrmion along the direction perpendicular to the in-plane component of the magnetic field. They also observed the distortion of the triangular skyrmion lattice into a centered rectangular lattice when a tilted field was applied. Leonov and K\'ezsm\'arki \cite{Leonov2017Tilted} extended this study to N\'{e}el-type skyrmions with the uniaxial anisotropy. Masaki \textit{et al} \cite{Masaki2018} generalized these results to chiral solitions in monoaxial chiral magnets and saw the interaction between solitions goes from repulsion to attraction depending on their separation and the strength of the in-plane component of the applied field. Osorio \textit{et al} \cite{Osorio2019} studied skyrmion stability in ferromagnetic chiral magnets by studying the deformation of the skyrmion shape via harmonic expansion series of the skyrmion's boundary to see the effect of different fields. Wan \textit{et al} \cite{Wan2019} studied skrymion crystal stability in thin film helimagnets in a tilted field and observed that both the strength and angle of the field affected the skyrmion crystal stability. Kuchkin and Kiselev \cite{Kuchkin2020} studied 2D chiral skyrmions in a tilted field and saw the direction of the field and its angle can switch the skyrmion polarity and vorticity. They also observed the skyrmion-skyrmion attraction for certain tilted fields. 

The stability of skyrmions in the oblique field has been observed experimentally. S. Zhang \textit{et al} \cite{ZhangTilted2018} reported the evolution of skyrmions into snake-like structures dominated by chirality on application of the oblique field. Bord\'acs \textit{et al}. \cite{Bordacs2017} saw that the N\'{e}el-type skyrmion lattice is stable even with the field tilted $54.7^{o}$ from the polar axis in the easy plane magnet $\mathrm{GaV_4Se_8}$.  Expanding on this, Gross \textit{et al}. \cite{Gross2020} determined the stability of the skyrmion lattice in $\mathrm{GaV_4Se_8}$, as well as $\mathrm{GaV_4S_8}$, as a function of the tilt angle and field magnitude. Wang \textit{et al}. \cite{Wang2017} observed a skyrmion lattice in an oblique field at temperatures near the boundary between the conical and skyrmion phase of FeGe. Ding \textit{et al}. \cite{Ding2020} saw that a N\'{e}el-type skyrmion bubble lattice is stable in the material  $\mathrm{FeGeTe_2}$ in a tilted field. Du \textit{et al}. \cite{Du2018} saw that the nature of the skyrmion-boundary interaction depends on the strength of the field, in a study including a tilted field, in a cubic chiral magnet.

In this paper, we propose a different kind of a binary or quaternary memory device that involves the manipulation of the skyrmion between two or four energy minima where the oblique field modifies the repulsive interaction of the skyrmion with the boundary so that energy minima arise near some boundaries. The device we propose consists of a square area of a magnetic film of a finite thickness. We include all relevant magnetic interactions, using the effective 2D model which assumes an effective dipole-dipole interaction between the columns of spins comprising each atomic layer. The paper is organized as follows. The lattice model employed in the paper is introduced in Section \ref{Fixed_Spins} and solutions are obtained for stable skyrmion states in a square lattice in the oblique field. Relaxation of skyrmions towards the locations of the energy minima and their motion between the minima are studied in Section \ref{dynamics}. Realization of the binary and quaternary memory is studied in Section \ref{Devices}. Our conclusions are summarized in Section \ref{Sec_Conclusion}.

\begin{figure}[ht]
\hspace{-0.5cm}
\centering
\includegraphics[width=8cm]{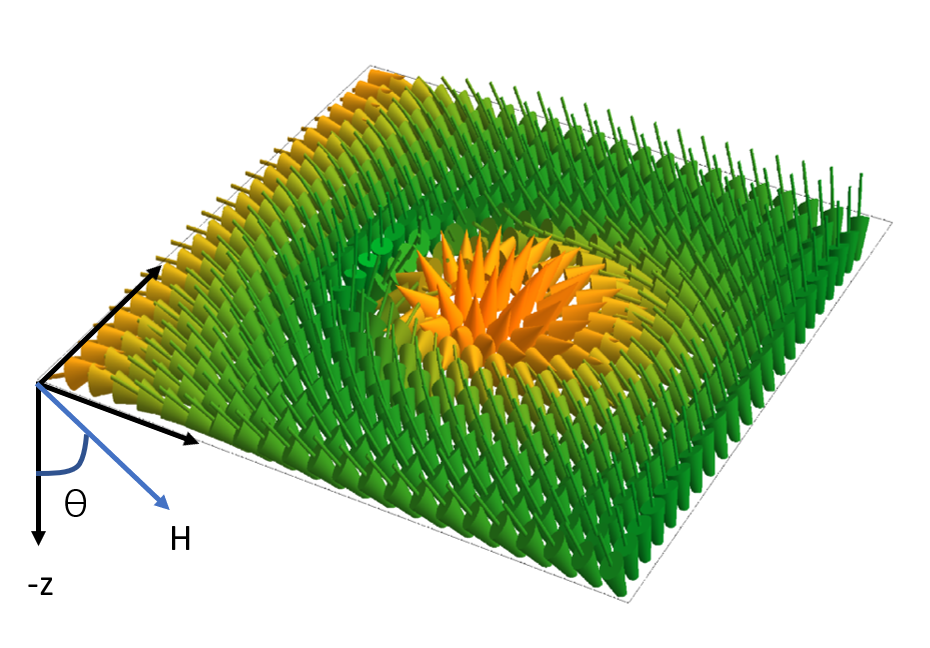} \includegraphics[width=8cm]{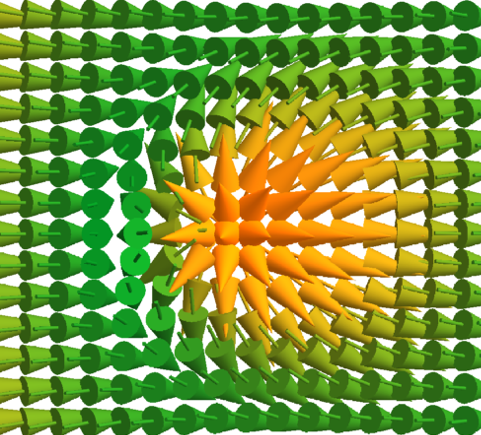} 
\caption{ N\'{e}el-type skyrmion of chirality $\gamma=0$ with fixed central spin $s_z=1$ in an oblique field with in-plane component along the \textit{x}-direction. We take the spins pointing up in the center of the skyrmion which means that the perpendicular component of the oblique field must be negative. Consequently, $\theta$ is measured with respect to the negative \textit{z}-axis, see upper panel. The lower panel is a zoomed-in region showing the deformation of the skyrmion along the direction of the oblique field.  }
\label{Skyrmion}
\end{figure}

\section{Skyrmion-Boundary Interactions in the Fixed Center-Spin Model} \label{Fixed_Spins}

In the numerical work, we consider a lattice model for a 2D ferromagnetic film  with the energy:
\begin{eqnarray}
{\cal H} & = &- \frac{1}{2} \sum_{i j}J_{ij}\mathbf{s}_{i}\cdot \mathbf{s}_{j}-\sum_{i } \mathbf{H} \cdot \mathbf{s}_{i} \nonumber \\
& - &A\sum_{i} \left[(\mathbf{s}_{i}\times \mathbf{s}_{i+\delta_{x}})_{y} - (\mathbf{s}_{i}\times \mathbf{s}_{i+\delta_{y}})_{x}\right] \nonumber \\ &-& \frac{D}{2}  \sum_{i} s_{iz}^2- \frac{E_{D}}{2}\sum_{ij}\Phi_{ij,\alpha\beta}s_{i\alpha}s_{j\beta}. \label{Hamiltonian}
\end{eqnarray}
The exchange coupling $J$ is for the nearest neighbors on a square lattice. The field:
\begin{equation}
\mathbf{H} = H \left(\cos\phi\sin\theta, \sin\phi \sin\theta,- \cos\theta \right), \label{H}
\end{equation}
has a perpendicular component $H_{z} = -H\cos\theta$ applied in the negative ($\theta<90^{\circ}$) \textit{z}-direction to prevent skyrmion expansion, with a magnitude smaller than the collapse field. Since we take the spins of the skyrmion pointing up at its center, the perpendicular component of the field must be downward and the angle $\theta$ is taken with respect to the negative \textit{z}-axis, see Fig. \ref{Skyrmion} . The in-plane field $H_{x}= H \sin\theta \cos\phi$ is produced when the net field is tilted away from this direction, i.e. $\theta \neq 0^{\circ}$. By adjusting the angle $\phi$ in the plane, a non-zero $H_{y}$ can be produced as well. Of note in this work is the angle $\phi=45^{\circ}$ corresponding to $H_{x}=H_{y}$. The DMI term favors outward  N\'{e}el-type skyrmions with in-plane components of chirality $\gamma=0$ for $A>0$. The subscripts  $\delta_{x}$ and $\delta_{y}$ refer to the next nearest lattice site in the positive $x$ or $y$ direction. Here $D$ is the easy-axis PMA constant and in the dipole-dipole interaction term (DDI):
\begin{equation}
\Phi_{ij,\alpha\beta}\equiv a^{3}r_{ij}^{-5}\left(3r_{ij,\alpha}r_{ij,\beta}-\delta_{\alpha\beta}r_{ij}^{2}\right),\label{DDIPhi}
\end{equation}
where $\mathbf{r}_{ij}\equiv\mathbf{r}_{i}-\mathbf{r}_{j}$ is the 
displacement vector between the lattice sites and $\alpha,\beta=x,y,z$ denote Cartesian components in a 3D coordinate space. The parameter $E_{D}$
defines the strength of the DDI. The relative strength of the PMA to the DDI is given by the dimensionless parameter $\beta \equiv D/(4\pi E_{D})$.  We start with one atomic layer where the energy is given by Eq. (\ref{Hamiltonian}) and use the definition of $\Phi$ from Eq.(\ref{DDIPhi}).

The effective 2D DDI model that we use for the film of $N_z$ layers assumes that the spin field does not depend on \textit{z}, and the effective DDI is between the columns of spins, in accordance with the methodology first introduced in Ref. [\onlinecite {Capic2019}]. Therefore, when the film has $N_z>1$ layers, Eq. (\ref{Hamiltonian}) gives the energy per layer. At large distances, the effective DDI divided by the number of layers is $N_z \Phi$, where $\Phi$ is defined in Eq. (\ref{DDIPhi}). 

This Hamiltonian is used to model a real material, specifically the PdFe/Ir(111) bilayer that has the  N\'{e}el-type DMI. Following the treatment in  
Ref. [\onlinecite{Leonov-NJP2016}], we extract the parameters $A/J=0.268$, $D/J=0.0236$ and $\beta=4.29$. This necessitates a field of strength $H/J=0.05$. The system was studied on square lattices of $300 \times 300$ and  $100 \times 100$ spins for the system with $N_z=1$ and $N_z=10$ layers. For the lattice spacing $a$ and the exchange constant $J$ we chose $a = J = 1$ in the computations.

The numerical method of the minimization of the energy  [\onlinecite{Garanin2013}] involves the rotation of the individual spins $\mathbf{s}_{i}$ towards the direction of the local effective field  $\mathbf{H}_{\mathrm{eff},i}= -\partial{\cal H} / \partial{\mathbf{s}_{i}}$, with the probability $\alpha$ and the energy-conserving spin flips (\textit{overrelaxation}), $\mathbf{s}_{i} \rightarrow 2(\mathbf{s}_{i} \cdot \mathbf{H}_{\mathrm{eff}, i})\mathbf{H}_{\mathrm{eff}, i}/H_{\mathrm{eff}, i}^2 -\mathbf{s}_{i}$ with the probability $1-\alpha$. The parameter $\alpha$ plays the role of the effective relaxation constant. We use the value $\alpha=0.03$ for the overall fastest convergence.

As the first step, we observed the well known skyrmion repulsion as the skyrmion approached the boundary for the field $H$ pointing completely downward for a single atomic layer $N_z=1$. Following the methodology of [\onlinecite{Capic2020}], the central spin of the skyrmion, where $s_{z}=1$, was fixed by applying a strong fictitious field in the positive \textit{z}-direction. To observe the repulsion, the pinning field was then moved one lattice spacing closer to the boundary per step,  and the skyrmion relaxed to the minimum energy in accordance with our numerical routine. The energy $E$, comprising the exchange, Zeeman, DMI, PMA and DDI energies was recorded as a function of the distance from the boundary to the fixed central spin, given by $d/a$. When the skyrmion is moved to the right (left), this corresponds to moving the pinning field one lattice spacing to the right (left) at fixed $n_{y}/a$. Similarly, when the skyrmion is moved to the top (bottom) of the lattice, the pinning field is moved one lattice spacing above (below) at fixed $n_{x}/a$. In this way, and since the skyrmion is small in comparison to the size of the lattice, we can determine the effect of a single boundary on the skyrmion. The behavior of the energy, which indicates repulsion with the boundary, is identical as the skyrmion approaches each of the four boundaries. In summary we see that for the field $\theta=0^{\circ}$, there is pure repulsion.

Next, still for one atomic layer $N_z=1$, we began to slowly increase the strength of the in-plane field from zero, which can physically be achieved by tilting the field so that $\theta>0^{\circ}$. To obtain an in-plane field in the positive \textit{x}-direction, one needs $\phi=0^{o}$.  Using the fixed central spin approach, a well defined minimum in the energy emerges with decreasing $d$ as the skyrmion approaches the top and bottom boundaries for $H_{x} \approx |H_{z}|/2$ for this set of parameters.

\begin{figure}[ht]
\hspace{-0.5cm}
\centering
\includegraphics[width=8cm]{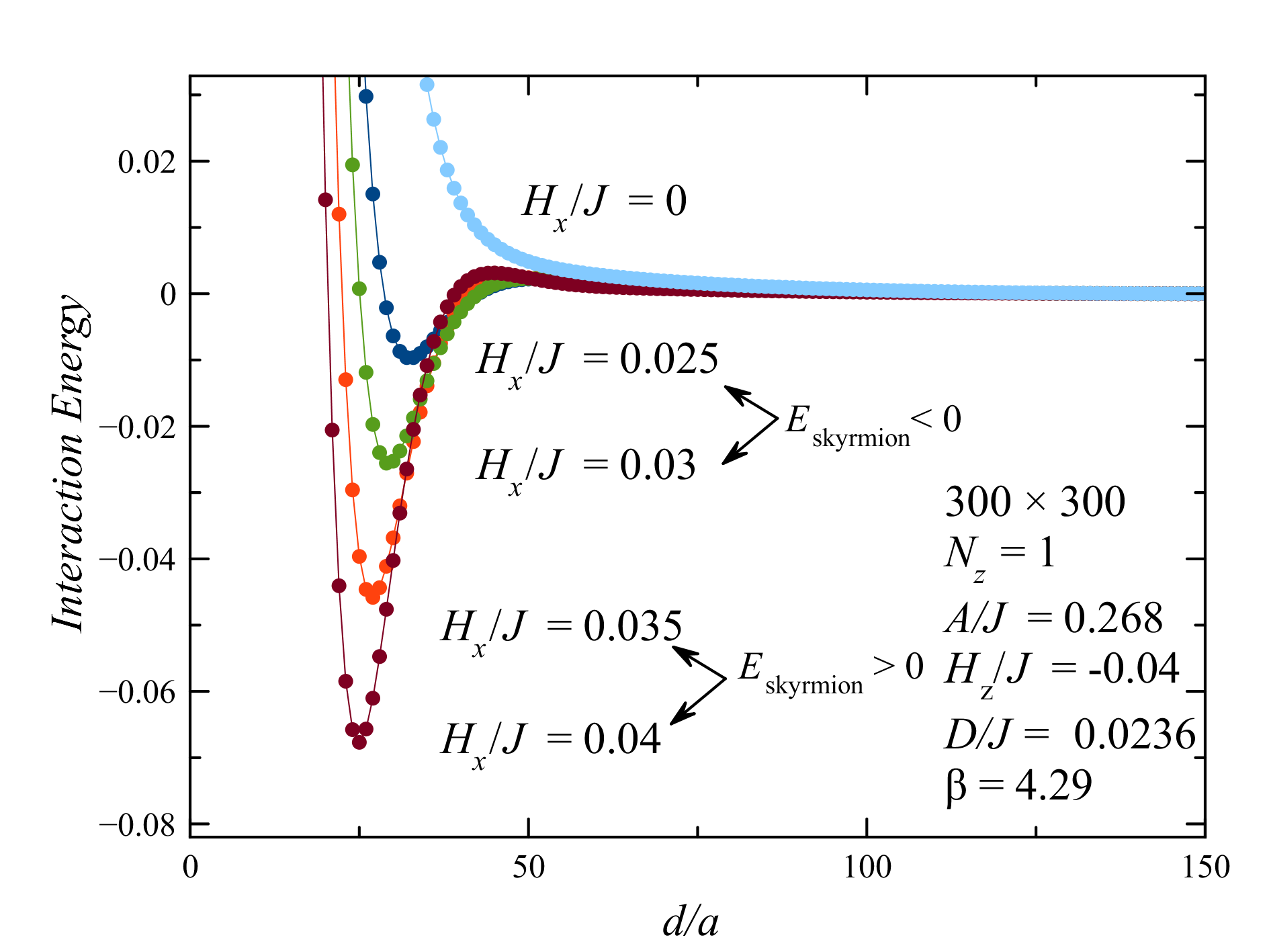}
\caption{The local minimum in the energy for the finite distance $d/a>0$ of the skyrmion center from the edge of the lattice gets deeper and the local maximum higher as the in-plane field strength $H_x$ is increased. The skyrmion is pinned at each $d$ using the fixed central spin. In this instance the behavior of the energy is identical as the skyrmion approaches the top and bottom boundaries at fixed $n_x/a$. There is only repulsion present as the skyrmion approaches the right and left boundaries at fixed $n_x/a$. Pictured here is  the path at fixed  $n_x/a=150$ for the skyrmion approaching the top boundary. $E_{skyrmion}$ denotes the energy of the skyrmion after subtracting the contribution from the uniform state. The interaction energy (the computation of which is discussed in the text) with the boundary is plotted for comparison purposes.}
\label{HxBeh}
\end{figure}

\begin{figure}[ht]
\hspace{-0.5cm}
\centering
\includegraphics[width=8cm]{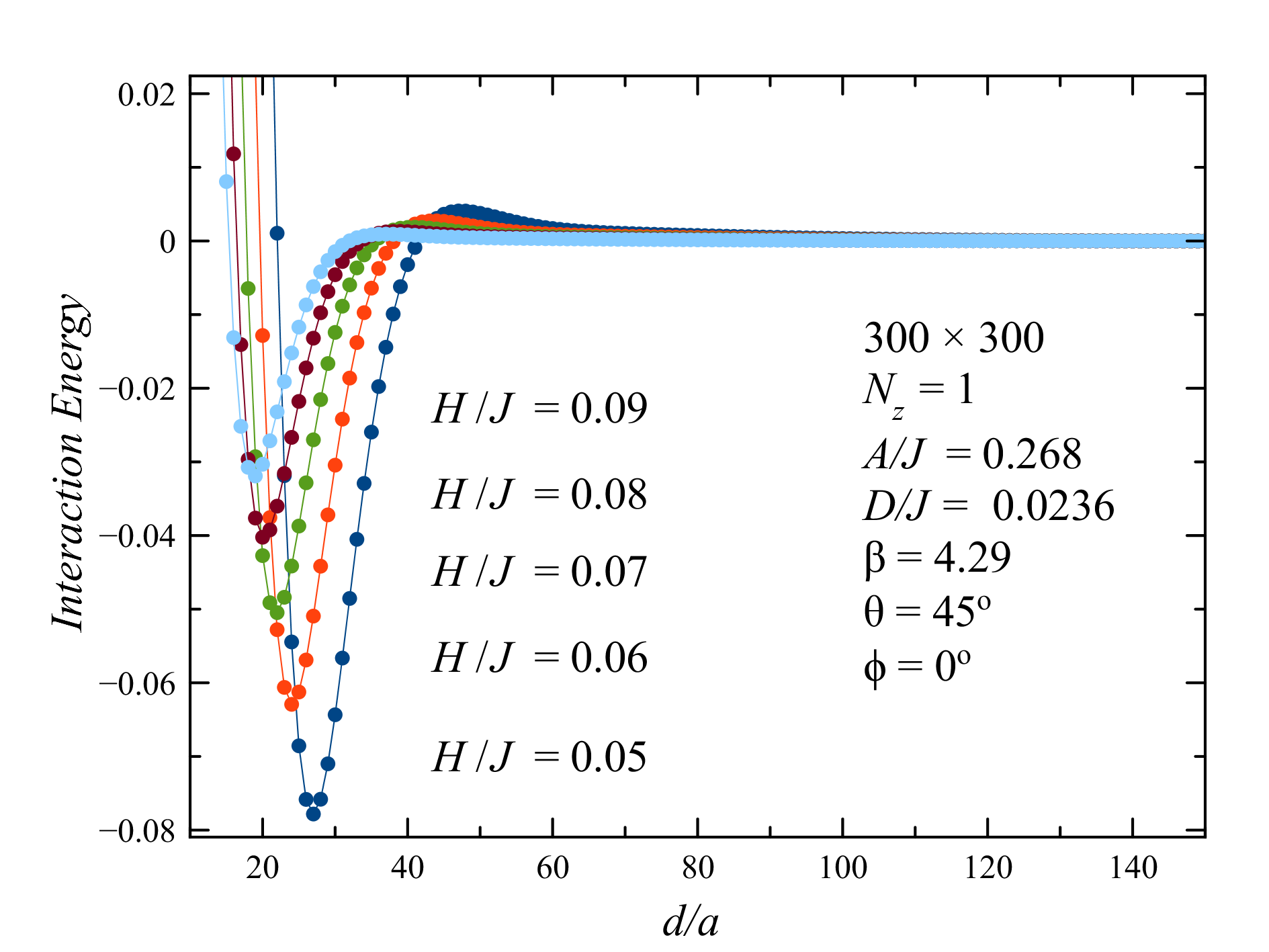}
\caption{Larger applied fields $H$ at fixed $\theta$ decrease the energy barrier. The energy is computed at each $d$, which is the distance of the central spin from the boundary, using the fixed central spin approach. Here, the behavior of the energy as the skyrmion approaches the top boundary at fixed $n_x/a=150$ is pictured.  The interaction energy (the computation of which is discussed in the text) with the boundary is plotted for comparison purposes.}
\label{DifferentHTwoMin}
\end{figure}

On increasing the in-plane field strength $H_x,$ for fixed $H_z,$  (physically one would have to vary both the total field strength $H$ and the angle $\theta$ to achieve this), the attractive potential that is present with decreasing $d$ as the skyrmion approaches the top and bottom boundaries also becomes deeper. Additionally, the energy minimum moves closer to the edge of the lattice with increasing $H_{x}$, i.e. the minimum is found at smaller $d$. Beyond a certain critical distance from the edge of the lattice, the repulsion dominates.  To compare the energies consistently, for Figs. \ref{HxBeh}-\ref{DifferentHTwoMin} we studied the interaction energy of the skyrmion with the boundary. This is found by computing the energy of the skyrmion at practically the center of the lattice with the fixed central spin pinned at $n_x/a=150,\; n_y/a=150$, and subtracting this from the energy of the skyrmion as $d$ decreases.  From Fig. \ref{HxBeh} , one can see that this behavior of the interaction energy is irrespective of whether the energy of the skyrmion itself is positive or negative compared to the uniform state.  Furthermore, stronger $H_{x}$ also leads to the formation of a local maximum in the energy as well.

On the contrary, the local minimum in the energy is shallower on increasing the magnitude of $H_z$ for fixed $H_x$ as the skyrmion approaches the top and bottom boundaries, i.e. with decreasing $d$. This effect is stronger than the effect of increasing $H_{x}$ for fixed $H_z$. One can see this by increasing the total field $H$ for fixed $\theta$ in which both components $H_{x}$ and $H_{z}$ are increased simultaneously, see Fig. \ref{DifferentHTwoMin} . One finds in this instance, like for the case of increasing $|H_z|$ for fixed $H_x,$ that the local minima are shallower with stronger total $H$.

The symmetry for the N\'{e}el-DMI is such that minima will form near the boundaries that are perpendicular to the direction of the in-plane field. In other words, for the present case where $\phi=0^{\circ}$, $H_{x}\neq 0$ and $H_{y}=0$, there will be a minimum in the skyrmion energy near the top and bottom boundaries and pure repulsion near the right and left boundaries. If one studied the Bloch DMI on the other hand, one would see the same qualitative behavior on the application of the fields just mentioned, only the boundaries where minima were present would be interchanged with the boundaries where pure repulsion were present for the N\'{e}el case.

\begin{figure}[ht]
\hspace{-0.5cm}
\centering
\includegraphics[width=8cm]{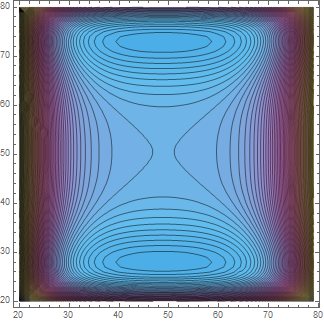}
\caption{The energy landscape as a contour plot for a $60 \times 60$ section of the $100\times 100$ lattice for $H/J=0.05$,  $\theta=45^{\circ}$, $\phi=0^{\circ}$ and $N_z=1$. The local minima are located near the top and bottom areas where the contours are blue. The purple regions towards the center, as well as the right and left edges are areas of higher energy. The darker regions beyond the purple areas are even higher energy. The energy in the regions closer to the boundaries that is not pictured is the highest.}
\label{EnerLandTwoMin}
\end{figure}

To obtain the energy landscape, we used the fixed central spin approach to pin the skyrmion at each lattice site and determined the energy using the already discussed energy minimization routine. The energy landscape was performed for a subsection of the $100\times100$ lattice, see Fig. \ref{EnerLandTwoMin}. The energy is lowest near the top and bottom boundaries, as already stated. Beyond these regions, the energy rises considerably. In particular, the energy towards the right and left boundaries is much higher than the top and bottom boundaries, indicating repulsion. Additionally, beyond the minima near the top and bottom boundaries, the energy also increases significantly, indicating repulsion as the skyrmion moves past these minima. 

Continuing the study for one atomic layer $N_z=1$, if the angle of the field $H$ in the plane is $\phi=45^{o}$, so that $H_{x}= H_{y}$, there are four minima. For the  N\'{e}el-DMI, the behavior of the energy as a function of the center of the skyrmion from the boundary using the fixed central spin approach is the same for the specific paths when the skyrmion approaches the right and top boundaries as well as the left and bottom boundaries, at fixed $n_y/a=150$ and $n_x/a=150$, respectively. In other words, there are two pairs of unequal minima near each of the four boundaries. Furthermore, the pairs of minima are closer in magnitude on increasing the strength of the field $H$ at fixed $\theta$, at the cost of decreasing the depth of the larger minima. 

\begin{figure}[ht]
\hspace{-0.5cm}
\centering
\includegraphics[width=8cm]{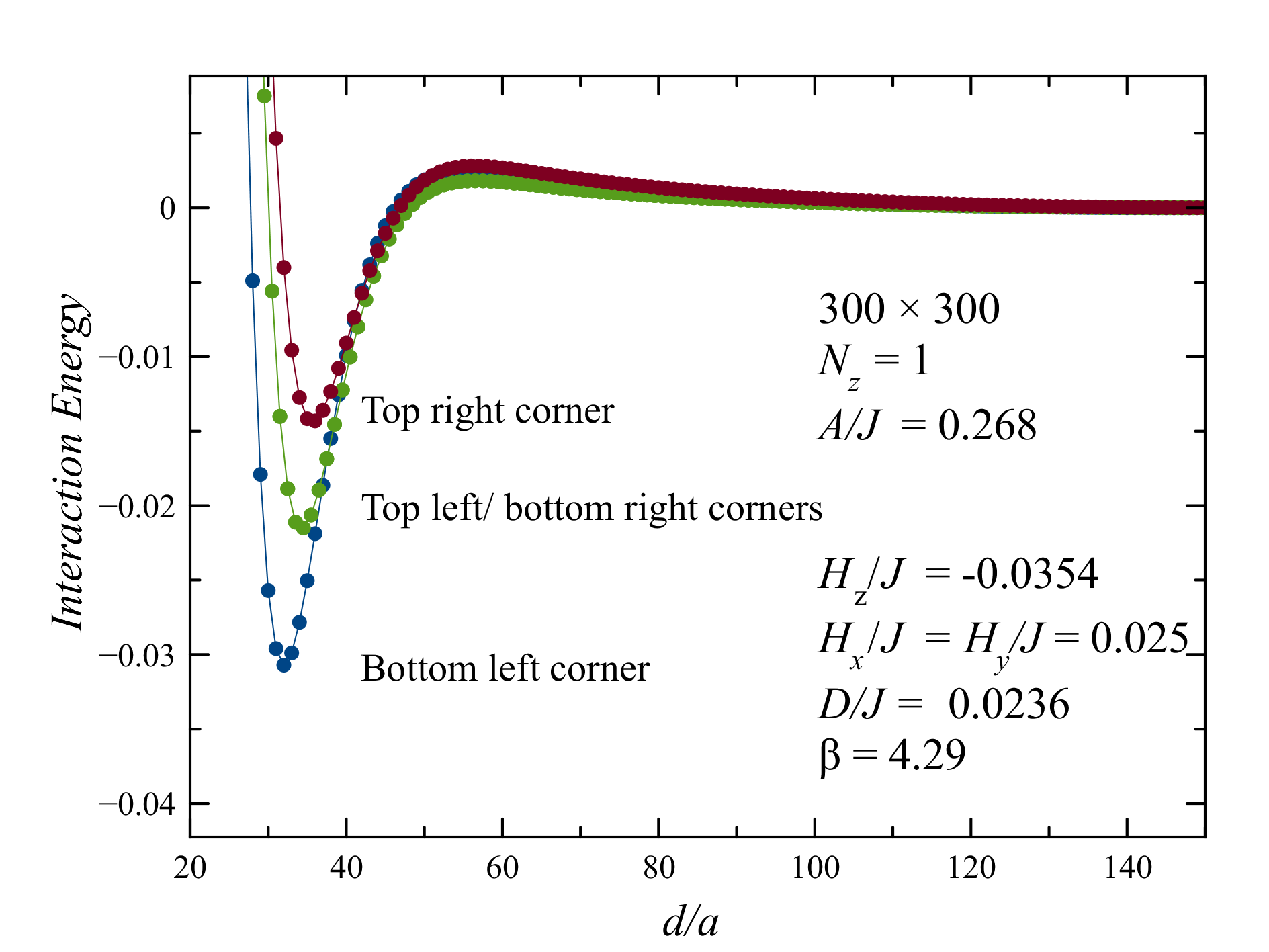}
\caption{Interaction energy of the skyrmion for the case $H_x=H_y$ using the fixed central spin approach, for the case of a single atomic layer, $N_z=1$. The pinning field pins the skyrmion along the diagonal path $n_x=n_y$ that goes through the bottom left and top right corners of the lattice, as well as the other diagonal path when $n_x= N-n_y$ that goes through the top left and bottom right corners of the lattice. $N$ is the length of one of the sides of the square lattice. The distance $d$ of the fixed central spin from the corner is decreased by moving the pinning field one lattice spacing horizontally and vertically per step. Here, $H/J=0.05$, $\theta=45^{\circ}$, $\phi=45^{\circ}$ and $N_z=1$.  The plotted $d$ is the distance from the corner divided by $\sqrt{2}$.}
\label{LeonovFourMinCorners}
\end{figure}

However, for this case the global minima are actually located near the corners of the lattice, along the diagonals, see Fig. \ref{LeonovFourMinCorners}. To obtain this data for the path of the skyrmion approaching the corners, the pinning field was moved one lattice spacing horizontally and vertically per step and the interaction energy with the boundary was obtained in the same way as Figs. \ref{HxBeh}-\ref{DifferentHTwoMin} .

\begin{figure}[ht]
\hspace{-0.5cm}
\centering
\includegraphics[width=8cm]{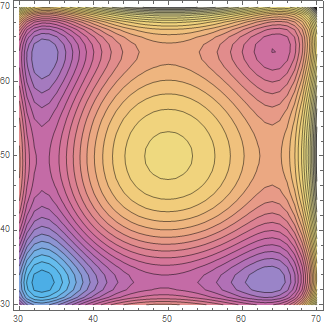}
\caption{The energy landscape as a contour plot for a $40 \times 40$ section of the $100\times 100$ lattice for $H/J=0.05$,  $\theta=45^{\circ}$, $\phi=45^{\circ}$ and $N_z=1$. The local minima are located along the diagonals where the contours are blue and purple. The deepest energy minimum located along one diagonal in the bottom left corner is light blue. Along the same diagonal in the top right corner is the shallowest energy minimum. There are also energy minima along the other diagonal near the top left and bottom right corners. The energy in the regions closer to the boundaries that is not pictured is much higher.}
\label{EnerLand}
\end{figure}

We explored this further by determining the energy landscape of a section of the $100\times 100$ lattice, see Fig. \ref{EnerLand}, such as was done in Fig. \ref{EnerLandTwoMin} . The introduction of the additional component of the in-plane field produces minima near the corners, along the lattice diagonals, as opposed to near the top and bottom boundaries near the center of the lattice along the \textit{x}-direction when only $H_x \neq 0$. The bottom left corner has the deepest energy minimum, while the top left and bottom right corners are energetically equivalent. The top right corner has the shallowest minimum. 

\begin{figure}[ht]
\hspace{-0.5cm}
\centering
\includegraphics[width=8cm]{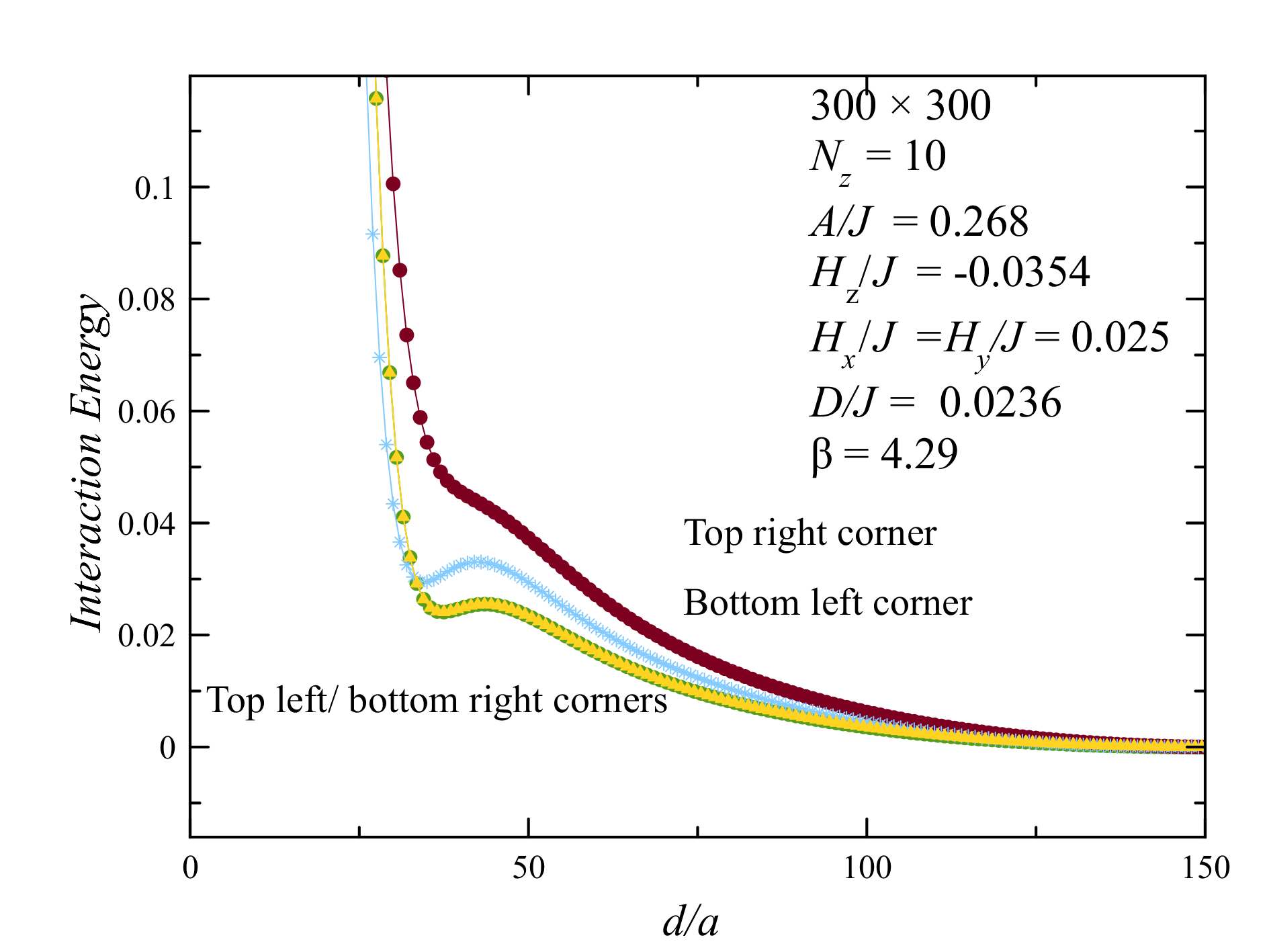}
\caption{Interaction energy of the skyrmion for the case $H_x=H_y$ using the fixed central spin approach for the case of $N_z=10$ atomic layers. The minimum near the top right corner vanishes with increasing number of atomic layers (compare to Fig. \ref{LeonovFourMinCorners} ).}
\label{LeonovFourMinNz10}
\end{figure}

Everything to this point was done for a single atomic layer, $N_z=1$. However, this work can be generalized to the system of arbitrary $N_z$ layers using the aforementioned effective 2D DDI between the columns of spins where we compute the energy per layer. Computations were also performed for the specific case $N_z=10$ layers.  

For the same parameters as the single layer case,  when $H_x\neq 0$ with $N_z=10$ layers,  the same qualitative behavior is observed in the energy, with minima near the top and bottom boundaries near the center of the lattice with respect to the \textit{x}-direction, and pure repulsion along the right and left boundaries. However, due to the long range behavior of the DDI, the local maximum in the energy is higher per layer and the local minimum is shallower per layer than the $N_z=1$ case. 

On the other hand, for the same parameters as the single layer case, for $H_x=H_y$ with $N_z=10$ layers, minima for the skyrmion approaching three of the four corners are positive with respect to the energy of the skyrmion computed in the center of the lattice, while the minimum when the skyrmion approaches the top right corner vanishes, see Fig. \ref{LeonovFourMinNz10} . Therefore, when the energy barrier is already anomalously small, (for these parameters it is $0.01J$),  it is advisable to increase the strength of the field along the directions in the plane, especially when increasing the number of atomic layers. For example, keeping the \textit{z}-component constant, one can increase the field strength to $H/J=0.0613$ and change the angle to $\theta=54.7354^{\circ}$ so that $|H_z|/J=H_x/J=H_y/J=0.0354$, leading to an energy barrier of $0.03J$ per layer.

\begin{figure}[ht]
\hspace{-0.5cm}
\centering
\includegraphics[width=8cm]{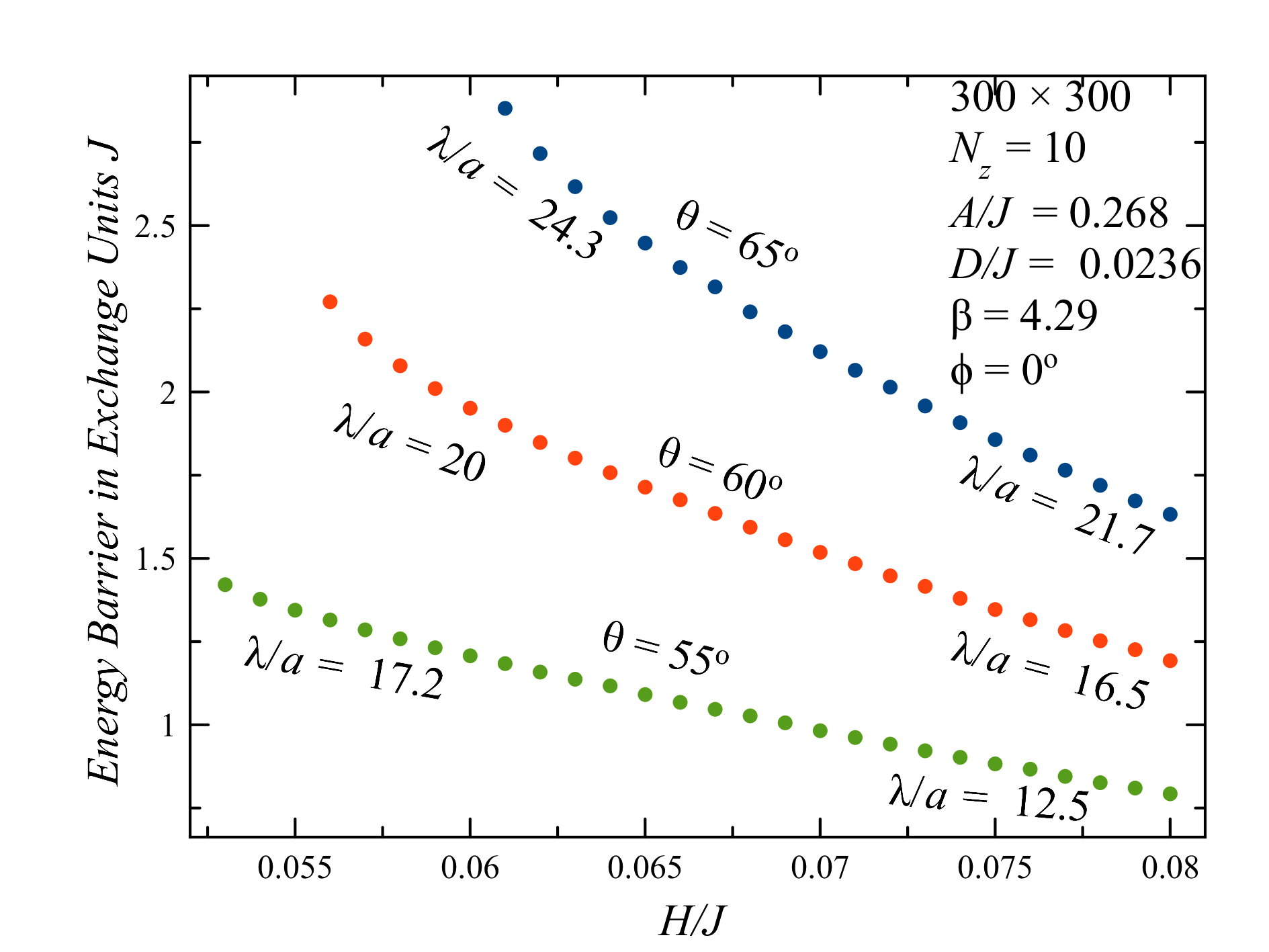}
\caption{The energy barrier in exchange units $J$ as a function of the applied field $H$ for several different angles $\theta$ and fixed $\phi = 0^{\circ}$ for $N_z=10$ layers. The room temperature energy barrier for this material  is $\approx 2J$. The barrier can be raised well above room temperature by using more layers.}
\label{Ebarr}
\end{figure}

The benefit of using more layers is that it is possible to increase the energy barrier, which is useful for potential room temperature applications, as the energy of the film scales with the number of layers $N_z$. The energy barrier is computed by determining the global minimum energy of the skyrmion using the aforementioned energy relaxation routine and then subtracting the energy found by fixing the skyrmion's central spin near the center of the lattice, $n_x/a=150$, $n_y/a=150$. The energy barrier, as was already seen, is increased with either increasing $H_x$ or decreasing $H$. Consequently, we observe that the barrier is highest with increasing angle $\theta$ when the component of the oblique field in the plane is larger than the perpendicular component. Stable skyrmions that have a room temperature energy barrier are only possible within a narrow range of $H$ and $\theta$ for $N_z=10$ layers, see Fig. \ref{Ebarr} . In general though, the barrier energy can be increased significantly higher than room temperature by using more layers, which would also require a smaller angle $\theta$.

\section{Skyrmion Dynamics in an Oblique Field}\label{dynamics}

To study the dynamics of the skyrmion in an oblique field, we solve numerically the system of Landau-Lifshitz equations for the spins on the lattice:
\begin{equation}
\dot{\mathbf{s}}_{i}=\frac{1}{\hbar}\mathbf{s}_{i}\times{\bf H}_{{\rm eff},i}-\frac{\alpha}{\hbar}\mathbf{s}_{i}\times\left(\mathbf{s}_{i}\times\mathbf{H}_{\mathrm{eff},i}\right), \label{LL}
\end{equation}
where $\alpha \ll 1$ is the damping constant, not to be misconstrued with the relaxation constant of the previous section. Fourth-order Runge-Kutta ordinary-differential-equation solver with the integration step $0.2$ in the units of $\hbar/J$ has been used. When the interactions are weak, the dynamics is rather slow, so that the discretization error of the Runge-Kutta method is rather small. However, one cannot increase the integration step past 0.25 since at about the step value of 0.3 an instability develops. Computation was performed on a $100 \times 100$ lattice using $\alpha = 0.1$ and free boundary conditions (fbc). Wolfram Mathematica with vectorization and compilation has been used on a 20-core Dell Precision Workstation (with 16 cores utilized by Mathematica).

\begin{figure}[ht]
\hspace{-0.5cm}
\centering
\includegraphics[width=8cm]{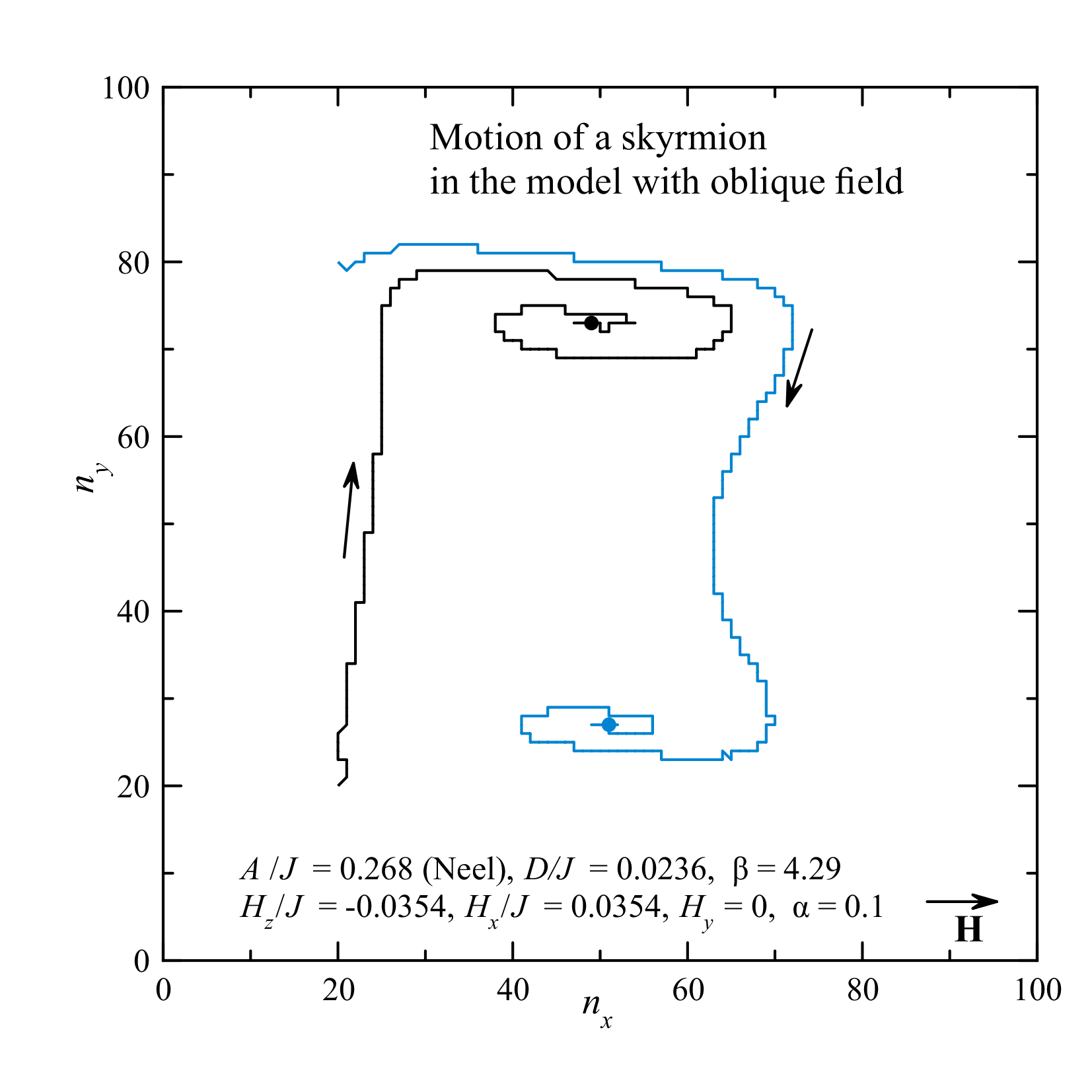}
\caption{The trajectory of the skyrmion when $H_x\neq 0$. The skyrmion will spiral about the minima located towards the top and bottom of the lattice for the case $H/J=0.05$, $\theta=45^{\circ}$ and $\phi=0^{\circ}$.}
\label{SpiralHx}
\end{figure}

We began by studying the system with one component of the field in the plane, $H_x\neq 0$ using the same parameters as the previous section. The numerical minimization of the energy using the fixed central spin method showed that for this situation, there are two minima in the energy located near the top and bottom boundaries of the lattice near the center of the lattice with respect to the \textit{x}-direction. We see from the dynamics that the skyrmion will spiral about these minima before eventually settling at them, see Fig. \ref{SpiralHx} . The skyrmion trajectory is tracked by identifying the position of the center of the skyrmion where $s_z=1$. The trajectory is practically the same for $N_z=1$ or $N_z=10$ atomic layers. This is computed numerically by finding where the \textit{z}-component of all the spins is at a maximum.

\begin{figure}[ht]
\hspace{-0.5cm}
\centering
\includegraphics[width=8cm]{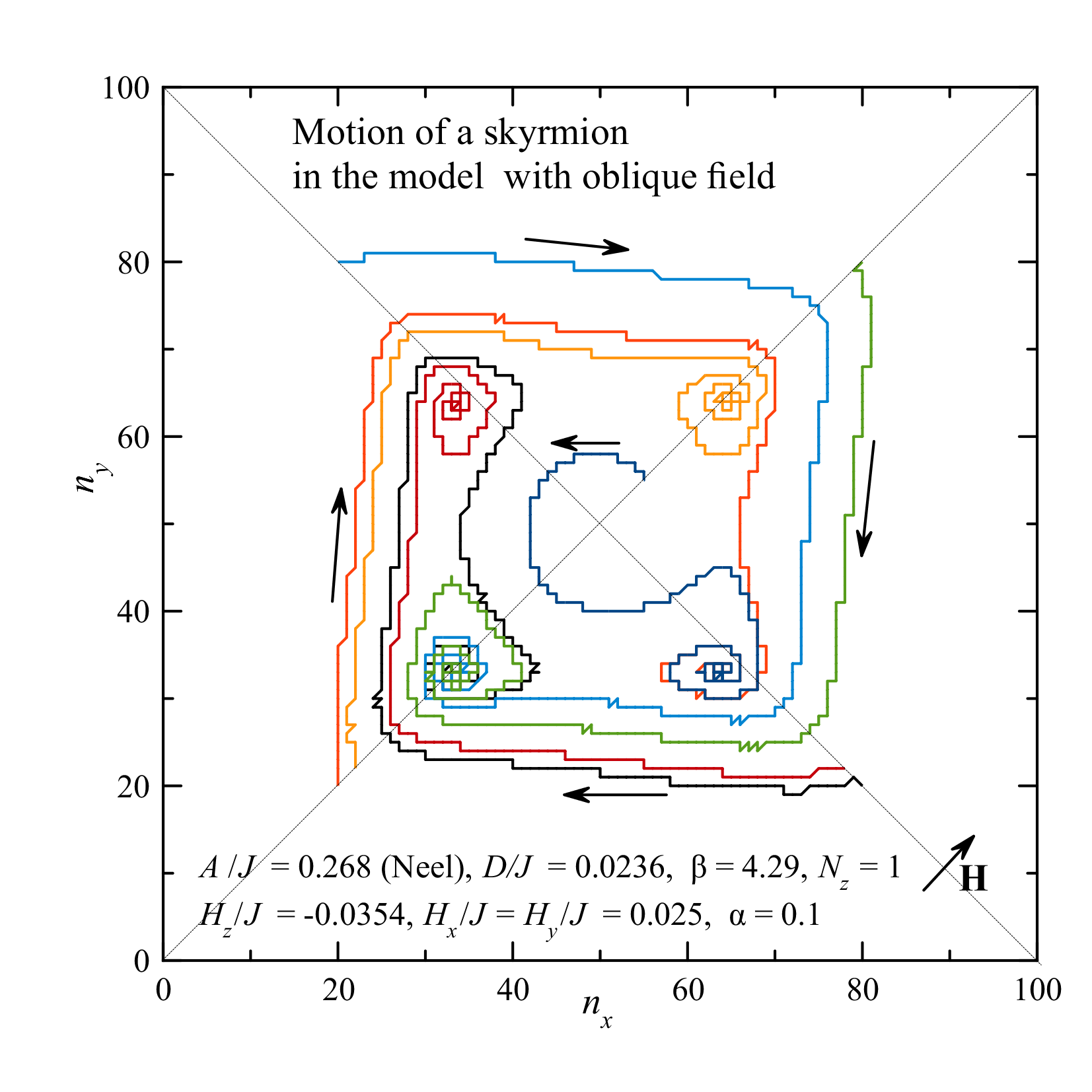}
\caption{The trajectory of the skyrmion when $H_{x}=H_{y}$ and $N_z=1$.  The minima are located towards the corners of the lattice, as already observed in the previous section. The skyrmion will spiral towards these minima. }
\label{SpiralCorners}
\end{figure}

For the case $H_x=H_y$, using the same parameters as the previous section, the energy minima are near the corners of the lattice, along the diagonals when there is one atomic layer, $N_z=1$. In this instance, the dynamics show that the skyrmion will also spiral towards these minima, see Fig. \ref{SpiralCorners} . 

\begin{figure}[ht]
\hspace{-0.5cm}
\centering
\includegraphics[width=8cm]{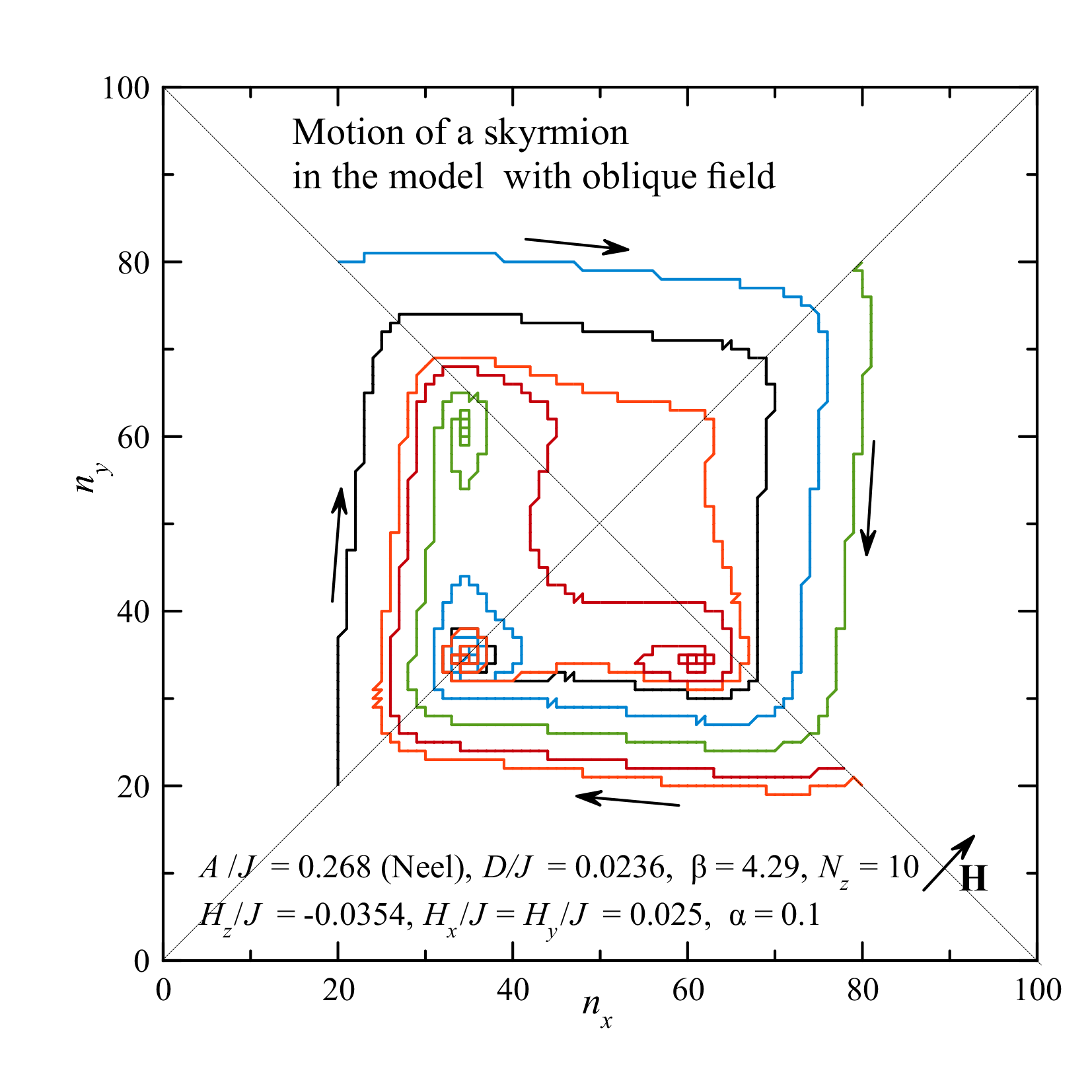}
\caption{The trajectory of the skyrmion when $H_{x}=H_{y}$ for $N_z=10$ layers. Even though the DDI is relatively weak, when the energy barrier is already anomalously small, as in the  $N_z=1$ case for these parameters, the energy minimum located near the top right corner will vanish with increasing number of layers, and instead the skyrmion will spiral around the energy minima located near the other three corners of the lattice. }
\label{SpiralCornersNz10}
\end{figure}

In the previous section, the numerical minimization of the energy revealed that the DDI has a negligible effect on the energy valleys for the case $H_x\neq 0$. However, for the case $H_x=H_y$ where the energy minima were already anomalously shallow,  when the number of layers is increased from $N_z=1$ to $N_z=10$, the minimum that was initially present near the top right corner of the lattice vanishes. This will also affect the skyrmion's trajectory, see Fig. \ref{SpiralCornersNz10} .

\section{Binary and Quaternary Memory}\label{Devices}

Thus we see that the application of an oblique field makes it energetically preferable for the skyrmion to settle a finite distance from the edge of the boundary in a  real material. For the oblique field with only one component in the plane, (in the above $H_{x}\neq 0$), these minima are near the top and bottom boundaries of the lattice. To use this as a potential binary skyrmion computation device, it is desirable to see if a magnetic gradient can be used to transfer the skyrmion in between these two minima.

\begin{figure}[H]
\hspace{-0.5cm}
\centering
\includegraphics[width=8cm]{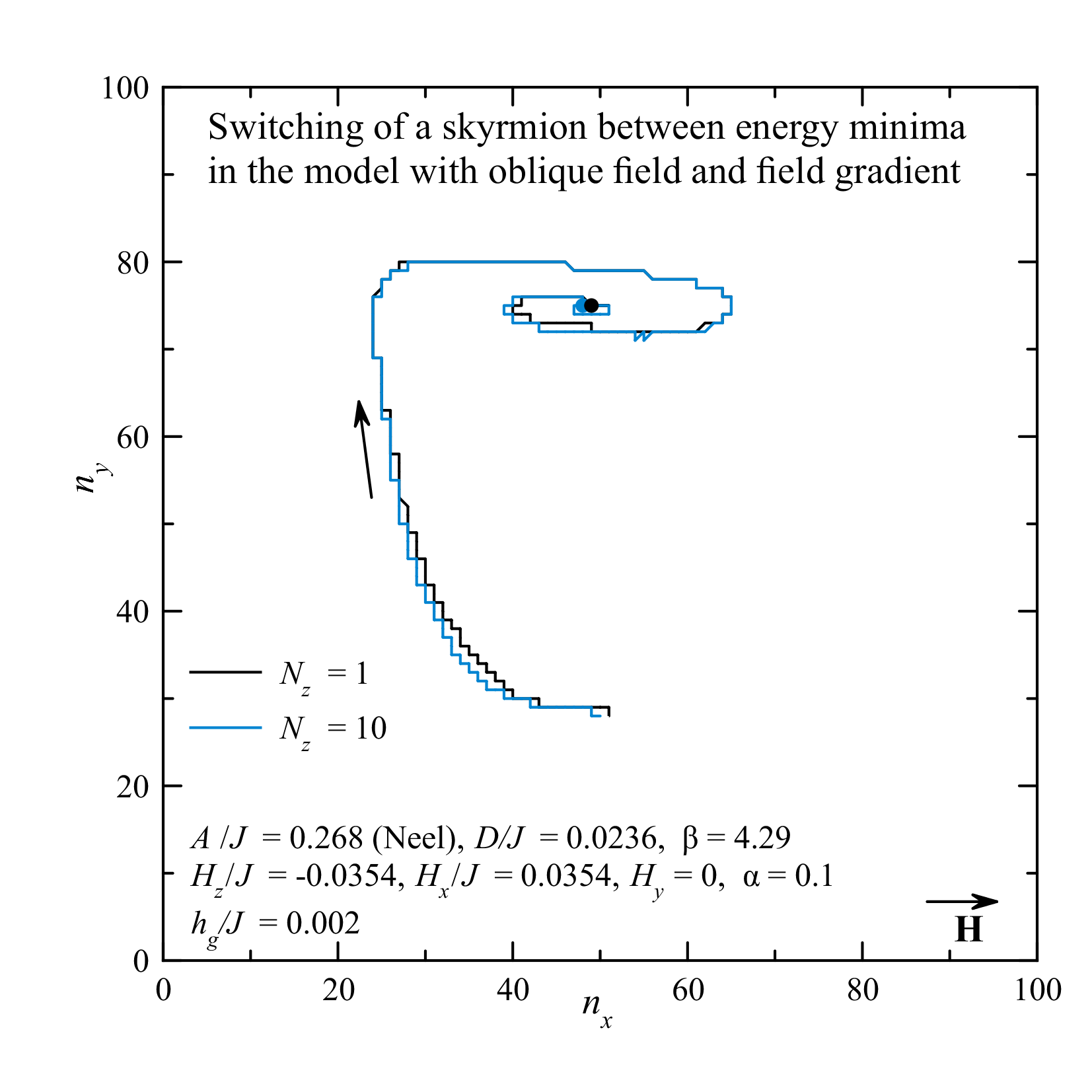}
\caption{The trajectory of the skyrmion when the gradient is applied for $H_x\neq 0$ for the system with $N_z=1$ and  $N_z = 10$ layers. The skyrmion can be moved from the minimum in the bottom boundary towards the minimum at the top boundary.  }
\label{SkyrmGradHx}
\end{figure}

To test this, we follow the same procedure of the numerical solution of the Landau-Lifshitz equation Eq. \ref{LL} with the additional term $-\sum_{i} \mathbf{H}_{g} \cdot \mathbf{s}_{i}$ included in Eq. (\ref{Hamiltonian}),  and without the fixed central spin. The skyrmion position is computed using the same method as the previous section. If the gradient, directed along the \textit{z}-direction has the form
\begin{equation}
H_g(n_x, n_y)_z = g \left(2 \frac{n_y}{N_y} -1\right), 
\end{equation}
where $g/J=0.002$ is the gradient field amplitude, and $N_y$ is the width of the lattice, then the skyrmion will switch between the bottom energy valley and the top energy valley. As before, the dynamics were performed on a $100\times 100$ lattice. During the course of the dynamics, the skyrmion will switch between the bottom valley at position $(50,28)$ to the top valley at position $(50,72)$ for $H/J=0.05$, $\theta=45^{\circ}$ and $\phi=0^{\circ}$. This behavior is practically the same regardless of whether there are $N_z=1$ or $N_z=10$ layers, see Fig. \ref{SkyrmGradHx} .

\section{Conclusions} \label{Sec_Conclusion}

Skyrmion states and dynamics in a square island in the oblique field have been studied. Using as an example the parameters of PdFe/Ir(111) and accounting for all relevant interactions, we have demonstrated within an effective 2D lattice model that an oblique field induces two or four spatially separated stable states of a skyrmion in a square island.  

The field tilted towards the side of the square generates two energy minima while the field tilted towards the diagonal of the square generates four minima. The depth of the energy minima and the barriers between them can be controlled by the strength and orientation of the field. For the diagonal in-plane field, the minima are much shallower than for the in-plane field along \textit{x}. The energy minima can be well above room temperature in the island containing a few tens of atomic layers. 

The dynamics of the skyrmion in the island consists of a spiral motion towards one of the energy minima. Skyrmions can be moved between any two energy minima via application of the field gradient chosen in a specific direction that has been worked out in the paper. A faster switching (not studied here) can be achieved with the help of a spin-polarized current. 

The proposed multiple, spatially separated, topologically protected, stable skyrmion states should not be difficult to implement in experiment. A room temperature device would generally consist of a square magnetic island, 100nm on a side and thickness of a few nanometers. Skyrmion states should be controlled by the strength and orientation of the magnetic field.  It will open a possibility of binary and quaternary computer memory based upon skyrmions.

\section{Acknowledgements}

This work has been supported by the grant No. DE-FG02-93ER45487 funded
by the U.S. Department of Energy, Office of Science.

\end{document}